\documentclass[12pt]{article}

\usepackage{graphicx}
\usepackage{amsmath}
\usepackage{amssymb}
\usepackage{xcolor}

\newcommand{\F}{\textsf{F}}
\newcommand{\GPDf}{\mathfrak{f}}
\newcommand{\GDA}{\mathrm{F}}
\newcommand{\GDAtH}{\widetilde{\mathrm{H}}}
\newcommand{\GDAf}{\mathrm{f}}
\newcommand{\GPDh}{\mathfrak{h}}
\newcommand{\GDAh}{\mathrm{h}}
\newcommand{\GPDth}{\widetilde{\mathfrak{h}}}
\newcommand{\GDAth}{\widetilde{\mathrm{h}}}

\begin{document}

\begin{titlepage}

\centerline{\large \bf Inverse Radon transform at work}

\vspace{7mm}

\centerline{\bf I.~R.~Gabdrakhmanov$^{a}$,  D.~M\"uller$^{b}$ and O.~V.~Teryaev$^{a,c}$}

\vspace{7mm}

 \centerline{\it $^a$ Veksler and Baldin Laboratory of High Energy Physics, JINR} \centerline{\it  141980 Dubna, Russia}

\vspace{4mm} \centerline{\it $^b$Institut f\"ur Theoretische
Physik II, Ruhr-Universit\"at Bochum} \centerline{\it 44780
Bochum, Germany}

\vspace{4mm}

 \centerline{\it $^c$ Bogoliubov Laboratory of Theoretical Physics, JINR} \centerline{\it  141980 Dubna, Russia}

\vspace{4mm}

\vspace{20mm}

\begin{abstract}
\noindent
The inverse Radon transform allows to obtain partonic double distributions from (extended) generalized parton distributions. We express
the extension of generalized parton distributions by their dual parts, generalized distribution amplitudes and study some aspects of the
filtered backprojection (inverse Radon transform). We also show that  single integral transforms, previously obtained in the context of
wave function overlap representation, are valid for generalized parton distributions that do not possess such a representation.
Utilizing Radyushkin`s double distribution ansatz, we study  and compare the numerical evaluation of double distributions within the filtered backprojection and single integral transforms along the imaginary and real axes.

\vspace{20mm}
Inverse Radon Transform, Generalized Parton Distributions, Generalized Distribution Amplitudes,  Double Distributions
\end{abstract}

\end{titlepage}

\section{Introduction}
\label{intro}

Generalized parton distributions (GPDs) were conceptually introduced in connection with the partonic description of
deeply virtual Compton scattering  \cite{Mueller:1998fv,Radyushkin:1996nd,Ji:1996nm} and
deeply virtual meson production  \cite{Radyushkin:1996ru,Collins:1996fb}.
In leading power w.r.t.~the inverse photon virtuality these processes factorize in a perturbatively calculable
hard-scattering part and universal, i.e., process-independent, however, conventionally defined GPDs \cite{Collins:1996fb,Collins:1998be}.
The broad interest on GPDs arises from the fact that they encode
non-perturbative dynamics of the constituents, i.e., partons  in
hadrons or even nuclei, on the amplitude level. In fact their first Mellin moments are related to the spin problem of the proton \cite{Ji:1996ek}, to gravitational couplings of quarks and gluons \cite{Teryaev:1999su,Teryaev:2016edw} and to the spatial distribution of pressure as well as shear forces \cite{Pol02}, recently, acquiring  the interdisciplinary dimension \cite{Burkert:2018bqq}.

Moreover, GPDs possess a probabilistic  interpretation \cite{Burkardt:2000za,Burkardt:2002hr}, and might
be represented as an overlap of light-front wave functions
\cite{Diehl:1998kh,Diehl:2000xz,Brodsky:2000xy}. Comprehensive reviews about GPDs, their interpretation, and the
phenomenology are given in Refs.~\cite{Diehl:2003ny,Belitsky:2005qn}.

On the other hand, they are intricate functions that have to satisfy both polynomiality and positivity \cite{Pire:1998nw} conditions. The former one is implemented in the so--called double distribution representation \cite{Mueller:1998fv,Radyushkin:1997ki}, which is nothing but a Radon transform, pointed out in Refs.\ \cite{Teryaev:2001qm,Belitsky:2000vk}, while the latter one is manifest in the wave function overlap representation. However, to our best knowledge there is only one rather cumbersome representation, given by a four--fold integral, that satisfy both conditions \cite{Pobylitsa:2002vi}. It is based  on a certain parametrization of the double distribution (DD).

The Radon transform combined with the dispersion relations implies \cite{Teryaev:2005uj,Anikin:2007yh} also the ``holographic"
property of GPDs when all the relevant information (at the leading order) about the GPDs in two-dimensional plane of partonic momentum fraction $x$ and the skewness $\eta$ is contained at the line $|x|=\eta$ and the subtraction constant, related to pressure \cite{Pol02} and used  for its experimental investigation \cite{Burkert:2018bqq}.

In connection with the problem to have a GPD parametrization at hand that satisfy all theoretical constraints, one might employ the wave function overlap representation, which provides us the GPD in the outer (DGLAP)  region, $|x|  > \eta$, and uses internal GPD duality \cite{Kumericki:2008di,Hwang:2007tb,Muller:2014tqa,Muller:2017wms}.
Thereby, one might also numerically utilize the inverse Radon transform in a discretized form to obtain the DD, which is then used to calculate the full GPD \cite{Chouika:2017dhe,Chouika:2017rzs}.

As we have seen DDs are equally important as GPDs. However, the inverse Radon transform is considered as ill--posed, i.e., the DD is not
a continuous function of the GPD.  Only a few studies of the inverse Radon transform are performed in the context of GPDs \cite{Gabdrakhmanov:2012aa,Muller:2017wms,Chouika:2017dhe,Chouika:2017rzs}.

The outline of the article is as follows. In Sec.\ \ref{Sec-Preliminaries} we introduce the DD-representation and represent the extension of the GPD in terms of generalized distribution amplitudes (GDAs) generally introduced in \cite{Diehl:1998dk}. We discuss three different possibilities for the inverse Radon transform. In Sec.\ \ref{Sec-examples} we provide simple analytic examples for the inverse Radon transform within two--dimensional Fourier transform, filtered backprojection, and single integral transforms. In particular, this provides us some insights into the filtered backprojection. Furthermore, we use the (extended) GPDs from Radyushkin`s DD ansatz \cite{Radyushkin:2013hca} to study the fastness and robustness of the filtered backprojection and single integral transforms along the imaginary axis as well as the real axis.
Finally, in Sec.\ \ref{Sec-Conclusions} we summarize and give conclusions. One appendix is devoted to the (re)derivation of single integral inverse transforms.

\section{Preliminaries}
\label{Sec-Preliminaries}

\subsection{GPDs as Radon transform}

The DD representation of GPDs is for charge even GPDs $H$ and $E$ not uniquely defined. However, by means of a `gauge' transformation \cite{Teryaev:2001qm} one can always bring it into a `standard' form
\begin{eqnarray}
\label{GPDF-DD}
F(x,\eta,t)= \int_{-1}^1\!dy\!\!\int_{-1+|y|}^{1-|y|}\!dz\, \delta(x-y-\eta\, z) f(y,z,t) + D_F(x,\eta,t)
\,,
\end{eqnarray}
where the DD is symmetric in $z$, i.e., $f(y,z,t)=f(y,-z,t)$. The Mellin moments w.r.t.\ $x$ are  expressed by even polynomials in $\eta$.
Thereby, the so-called $D$--term on the r.h.s.\ of (\ref{GPDF-DD}) \cite{Polyakov:1999gs},
\begin{eqnarray}
\label{GPDF-D}
D_F(x,\eta,t) = \theta(1-|x/\eta|) {\rm sign}(\eta) d_F(x/\eta)\,,
\end{eqnarray}
completes polynomiality.
It is expressed by an antisymmetric function $d_F(x)$, i.e., $d_F(x)=-d_F(-x)$. From (\ref{GPDF-DD}) and (\ref{GPDF-D}) follows that this $D$--term might be extracted in the limit
\begin{eqnarray}
\label{d_F(x)-1}
d_F(x)= \lim_{\eta \to \infty} F(x \eta,\eta,t) \quad \mbox{for} \quad  |x| \le 1\,.
\end{eqnarray}

Without loss of generality we consider in the following a GPD  for which a possible $D$--term is subtracted and the DD is restricted to non-negative  $y$ values. Such a GPD is given by a common Radon transform
\begin{equation}
\label{F-D}
F(x,\eta,t)-D_F(x,\eta,t) \Rightarrow F(x,\eta,t)= \int_0^1\!\!dy\!\!\! \int^{1-y}_{-1+y}\!\! dz\, \delta(x-y-\eta\, z) f(y,z,t)
\end{equation}
that has the compact DD--support $0\le y \le 1$ and $|z| \le 1-y$. Thus, for $|\eta| \le 1$ the momentum fraction is restricted
to $-\eta \le x \le 1$ and the GPD,
\begin{eqnarray}
F(x,\eta,t) = \theta(x+\eta)  \GPDf(x,\eta,t) +  \theta(x-\eta) \GPDf(x,-\eta,t)\,,
\label{F(x,eta,t)}
\end{eqnarray}
can be expressed due to the  function
\begin{eqnarray}
\label{GPDf(x,eta,t}
\GPDf(x,\eta,t) = \frac{1}{\eta}\int_0^{\frac{x+\eta}{1+\eta}}\!dy\, f(y,(x-y)/\eta,t)\,.
\end{eqnarray}
This defining GPD function should vanish at $x=-\eta$, however, it is not necessarily analytic at the point $\eta=0$.
Here and in the following $\eta$ is considered to be {\em positive}.

The Radon transform (\ref{F-D}) also ensures that the GPD can be uniquely extended in the whole $(x,\eta)$-plane
(a consequence of the fact that it is the Fourier transform of an entire analytic function \cite{Mueller:1998fv,Mueller:2005ed}, see Sec.\ \ref{Sec-Fourier transform}).
For $|\eta|\ge 1$ we find the representation
\begin{subequations}
\label{F-ext}
\begin{eqnarray}
\label{F-support}
\F(x,\eta,t) = \Theta(x,\eta)\, \GPDf(x,\eta,t) +  \Theta(x,-\eta)\, \GPDf(x,-\eta,t)\,,
\end{eqnarray}
where the support restriction
\begin{eqnarray}
\Theta(x,\eta)\equiv{\rm sign}(1+\eta)\theta\!\left(\!\frac{x+\eta}{1+\eta}\!\right)\theta\!\left(\!\frac{1-x}{1+\eta}\!\right)
\end{eqnarray}
ensures that the polynomiality condition holds true for general $\eta$ values.
\end{subequations}

\begin{figure}[t]
\begin{center}
\includegraphics[width=6cm]{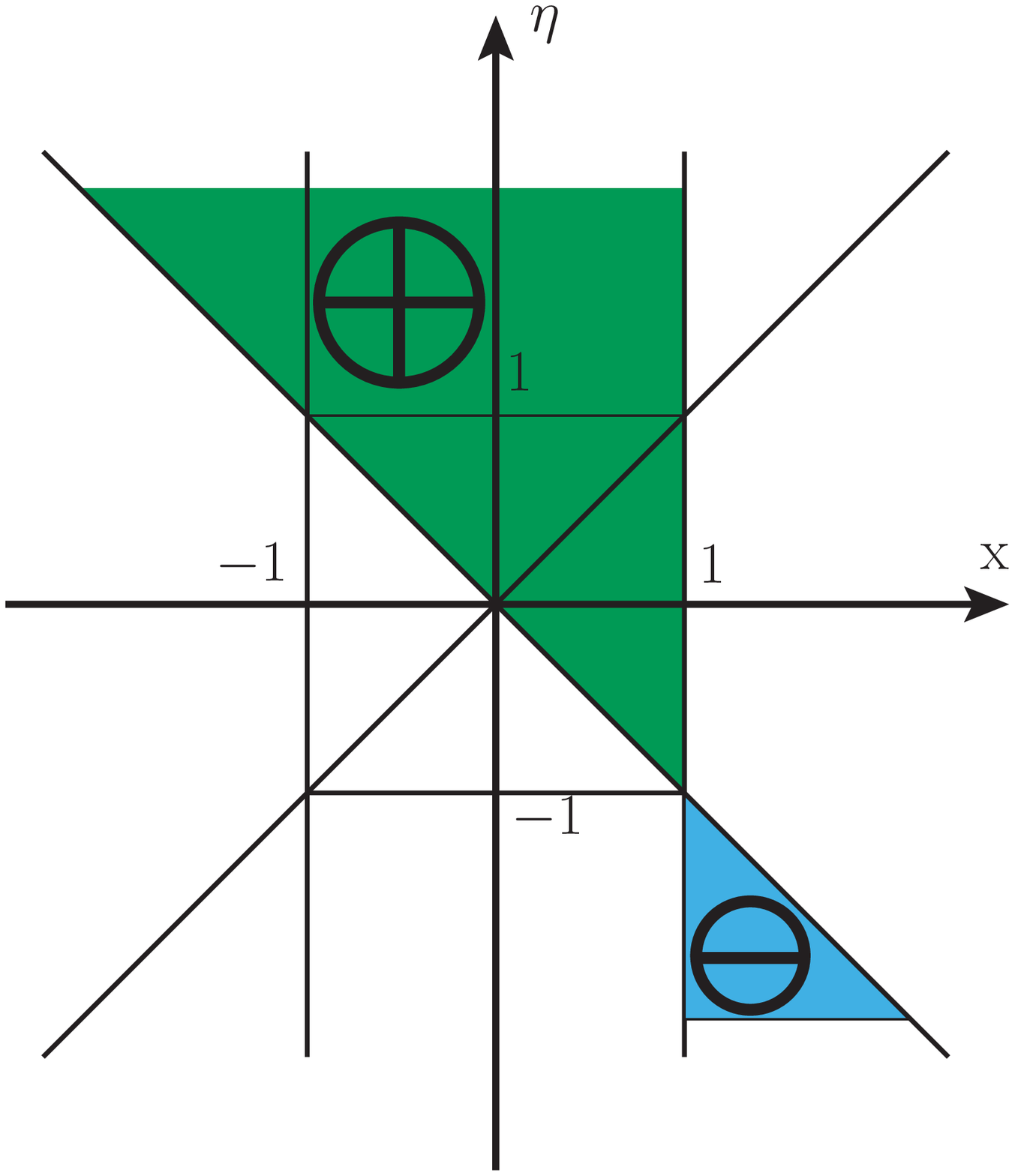}
\qquad
\includegraphics[width=6cm]{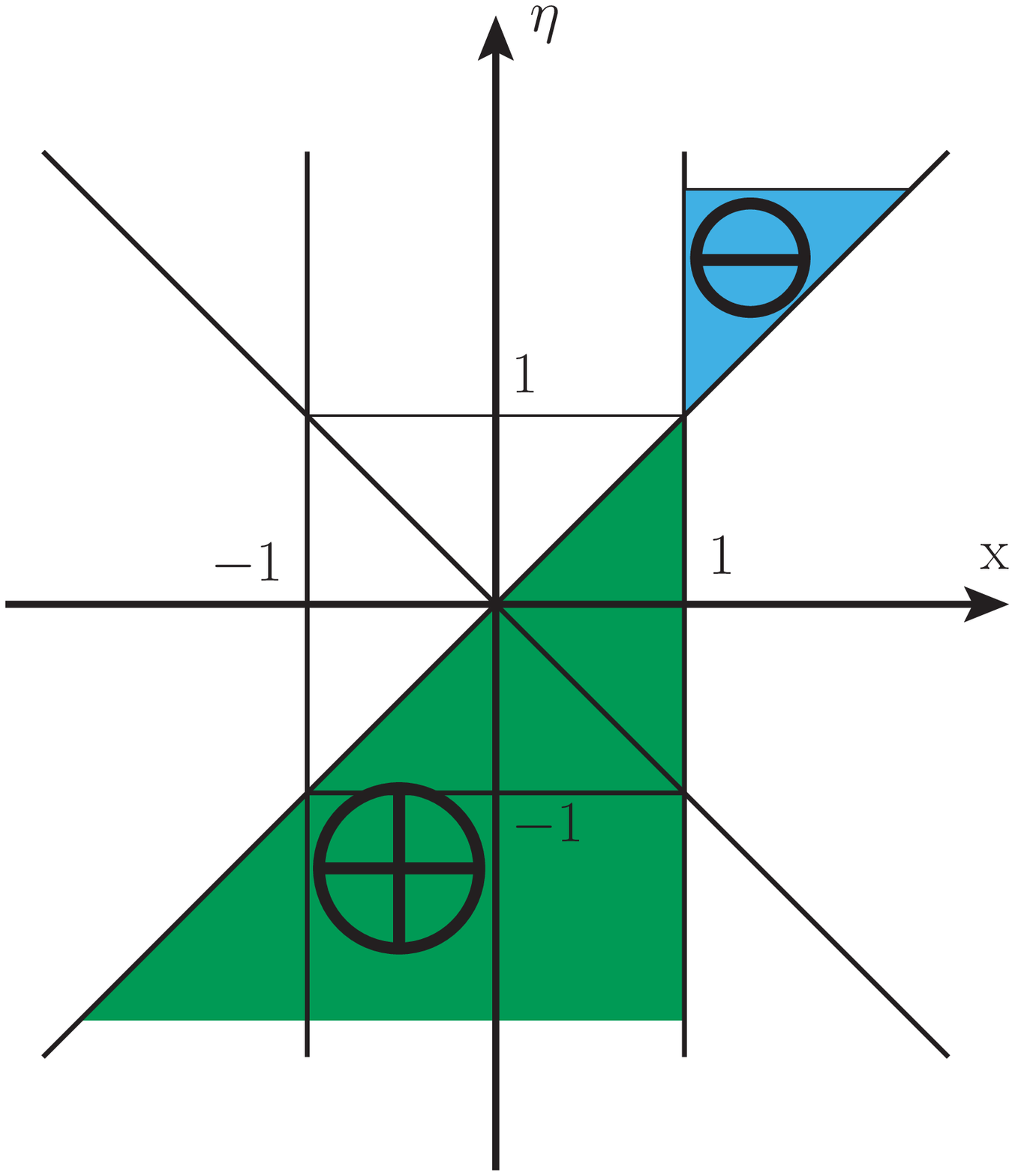}
 \end{center}
\caption{\small GPD support and its extension for the building blocks $\GPDf(x,\eta)$ (left) and $\GPDf(x,-\eta)$ (right),
where $\oplus$ and $\ominus$ denotes the pre--sign $+1$ and $-1$, respectively, of the extended functions.
\label{Fig-GPDsupport}}
\end{figure}
The support of the extended GPD (\ref{F-ext}), shown in Fig.\ \ref{Fig-GPDsupport}, can be explicitly written as
\begin{eqnarray}
\label{F-support1}
\F(x,\eta,t) &\!\!\!=\!\!\!& \theta(1-x) \theta(1-|\eta|) \left[ \theta(x+\eta) \GPDf(x,\eta,t) +\theta(x-\eta)  \GPDf(x,-\eta,t)\right]
\\
&&\!\!\!\!+ \theta(\eta-1) \left[\theta(1-x)\theta(x+\eta)\GPDf(x,\eta,t) -\theta(x-1)  \theta(\eta-x) \GPDf(x,-\eta,t)\right]
\nonumber\\
&&\!\!\!\!+\theta(-\eta-1)\left[ \theta(1-x)\theta(x-\eta) \GPDf(x,-\eta,t) -\theta(x-1) \theta(-\eta-x) \GPDf(x,\eta,t)\right].
\nonumber
\end{eqnarray}
It contains for $|\eta| \le 1$ the GPD part and for $|\eta| \ge 1$  two GDA  parts, which are related by reflection $\eta \to- \eta$,
\begin{eqnarray}
\F(x,\eta,t) &\!\!\!=\!\!\!& \theta(1-|\eta|) F(x,\eta,t) + \frac{\theta(\eta-1)}{\eta}   \GDA(x/\eta,1/\eta,t)+ \frac{\theta(-\eta-1)}{\eta}  \GDA(x/\eta,-1/\eta,t),
\end{eqnarray}
where the GDA is defined as
\begin{equation}
\GDA (x/\eta,1/\eta,t) = \eta\left[ \theta(1-x)\theta(x+\eta)\GPDf(x,\eta,t) -\theta(x-1)  \theta(\eta-x) \GPDf(x,-\eta,t)\right].
\end{equation}
Replacing $x\to x \eta$ and utilizing an inversion $\eta \to 1/\eta$ the GDA part reads for positive $\eta$
\begin{equation}
\GDA (x,\eta) = \frac{1}{\eta}\left[ \theta(\eta-x)\theta(x+1)\GPDf(x/\eta,1/\eta,t) -\theta(1-x) \theta(x-\eta)  \GPDf(x/\eta,-1/\eta,t)\right]
\end{equation}
or with $\GDAf(x,\eta) = 1/\eta \GPDf(x/\eta,1/\eta,t)$, i.e., also $\GDAf(-x,-\eta) = -1/\eta \GPDf(x/\eta,-1/\eta,t)$, we get
\begin{equation}
\GDA(x,\eta) =  \theta(\eta-x)\theta(x+1)\GDAf(x,\eta,t) + \theta(1-x) \theta(x-\eta)  \GDAf(-x,-\eta,t)\,.
\end{equation}

The GPD continuity is ensured by the boundary condition $\GPDf(x=-\eta,\eta,t)=0$, which corresponds to vanishing of the GDA at the endpoints, i.e., $\GDA (x=\pm 1,\eta)=0$. Vanishing of the quark GPD at $x=1$, i.e., $\GPDf(x=1,\eta,t)=- \GPDf(x=1,-\eta,t)$, ensures continuity of the GDA at $x=\eta$, i.e., $\GDAf(x=\eta,\eta) = \GDAf(x=-\eta,-\eta)$. Note also the symmetry relations
\begin{eqnarray}
F(x,\eta) = F(x,-\eta) \Leftrightarrow \GDA (x,\eta) = \GDA (-x,-\eta)
\end{eqnarray}
and that GPD and GDA are related to each other by reflection
\begin{eqnarray}
F(x,\eta,t)= \frac{1}{\eta}\GDA\!\left(\!\frac{x}{\eta},\frac{1}{\eta},t\!\right)  \Leftrightarrow \GDA(x,\eta,t)= \frac{1}{\eta}F\!\left(\!\frac{x}{\eta},\frac{1}{\eta},t\!\right),
\end{eqnarray}
i.e., the extension of the GPD/GDA support is given by the GDA/GPD. Alternatively to the limit procedure (\ref{d_F(x)-1}), the $D$--term might be obtained from the GDA
\begin{eqnarray}
\label{d_F(x)-2}
d_F(x)= \lim_{\eta \to 0} \eta \GDA(x,\eta,t) = \lim_{\eta \to 0} \eta \left[\theta(-x)\GDAf(x,\eta,t) + \theta(x)  \GDAf(-x,-\eta,t) \right]
\mbox{\ \ for \ \ }|x|\le 1.
\end{eqnarray}


\subsection{Inverse Radon transforms}

The inversion of the Radon transform (\ref{F-D}) is an ill--posed procedure in the sense that the resulting DD is not a continuous map of the GPD input.  Two inversion procedures are well--known: inversion by Fourier transform and the so--called filtered backprojection. In the latter case a Hilbert transform together with the derivative of the GPD serves in the language of signal processing as a filter. The backprojection is then done by a  single integral transform w.r.t.\ $\eta$. It turns out that if one works in the complex plane the filter is not needed and so only one integral remains, which we show in Appendix \ref{appendix}. This integral along the imaginary axis might  be also rewritten as an integral along the positive real axis, where the integrand is the imaginary part of the GPD in the central region. We are not aware that such single integral transforms are known in general.

\subsubsection{Inversion by Fourier transform}
\label{Sec-Fourier transform}

To evaluate the DD-function from a GPD, we perform first an inverse Fourier transform, where we can restrict ourselves to $|\eta|< 1$,
\begin{eqnarray}
\label{FTofGPD}
\mathfrak{F}^{-1}[F](\kappa,\eta,t) &\!\!\!=\!\!\!& \int_{-1}^1\!dx\, e^{- i \kappa x}F(x,\eta,t)\,.
\end{eqnarray}
This map provides us the inverse two--dimensional Fourier transform of the DD
\begin{eqnarray}
\mathfrak{F}^{-1}[F](\kappa,\eta,t) = \mathfrak{F}_2^{-1}[f](\kappa,\eta \kappa,t) \equiv\int_0^1\!\!dy\!\!\! \int^{1-y}_{-1+y}\!\! dz\, e^{- i \kappa y- i \kappa\eta\, z} f(y,z,t)\,,
\end{eqnarray}
which is an entire holomorphic function in both $\kappa$ and $\eta$. Thus, we can remove in principle the restriction on $|\eta|\le 1$ by analytic continuation (AC) w.r.t.\ $\eta$.
The Fourier transform provides us then the extended GPD
\begin{eqnarray}
\F(x,\eta,t) = \frac{1}{2\pi}\int_{-\infty}^\infty\!d\kappa\, e^{i \kappa x}  {\rm AC}\, \mathfrak{F}^{-1}[F](\kappa,\eta,t)\,.
\end{eqnarray}

The first method to obtain the DD is a two dimensional Fourier transform,
\begin{eqnarray}
\label{DDfromGPD1}
f(y,z,t) = \frac{1}{4\pi^2}\int_{-\infty}^\infty\!d\kappa\! \int_{-\infty}^\infty\!d\lambda\, e^{i \kappa y + i \lambda z}\, \mathfrak{F}^{-1}[\F](\kappa,\lambda/\kappa,t)\,,
\end{eqnarray}
of the inverse Fourier transform
\begin{eqnarray}
\label{invFTGPD}
\mathfrak{F}^{-1}[\F](\kappa,\eta,t) = \int_{-\infty}^\infty\!dx\, e^{- i \kappa x}\F(x,\eta,t)
\end{eqnarray}
of the extended GPD $\F$ or by means of
analytic continuation $\mathfrak{F}^{-1}[\F](\kappa,\eta,t) = {\rm AC}\, \mathfrak{F}^{-1}[F](\kappa,\eta,t) $ from  the GPD itself.

\subsubsection{Filtered backprojection formula}

Plugging the inverse Fourier transform (\ref{invFTGPD}) into the DD representation (\ref{DDfromGPD1}) yields as a second method for the
inverse Radon transform the filtered backprojection formula
\begin{eqnarray}
\label{F_i2f_i}
f(y,z,t)=\frac{-1}{2\pi^2}\!\! \int_{-\infty}^\infty\!\!\!d\eta\,\frac{\partial}{\partial y}  {\rm PV}\!\!\int_{-\infty}^\infty\!\!\frac{dx}{x}\;
\F(x+y+\eta z,\eta,t)
\,,
\end{eqnarray}
where the uniquely extended GPD  $\F$ is needed.  Employing the support properties of the extended GPD (\ref{F-support1}) and its symmetry w.r.t.~$\eta$, we can write the
inverse Radon transform after a shift of the $x$--integration variable, $x \to x-y-\eta z$, as
\begin{eqnarray}
\label{f(y,z,t|eta)-0}
f(y,z,t)&\!\!\!=\!\!\!&\int_0^\infty\!\!\!d\eta\, f(y,z,t|\eta)\,,
\\
\label{invRadon0}
f(y,z,t|\eta) &\!\!\!=\!\!\!& \frac{-1}{2\pi^2}\frac{\partial}{\partial y}\left[ {\rm PV}\!\!\int_{-\eta}^1\!dx \frac{2(x-y)}{(x-y)^2-\eta^2 z^2}\;
 \GPDf(x,\eta,t) + {\rm PV}\!\!\int_{\eta}^1\!dx \frac{2(x-y)}{(x-y)^2-\eta^2 z^2}\;
 \GPDf(x,-\eta,t) \right],
\nonumber\\
\end{eqnarray}
where we used the common rules for integrals, e.g., ${\rm sign}(1-\eta) \int_1^\eta\!dx\cdots = \int^1_\eta\! dx\cdots$ for $\eta >1$.
Furthermore, decomposing the $\eta$-integral into the two regions $[0,1]$ and $[1,\infty]$ and performing in the latter an inversion $\eta \to 1/\eta$
yields finally a two--component representation
\begin{eqnarray}
\label{f(y,z,t|eta)}
f(y,z,t)&\!\!\!=\!\!\!&\int_0^1\!\!\!d\eta\, f(y,z,t|\eta)\,, \quad f(y,z,t|\eta) = f_{\rm GPD}(y,z,t|\eta)+ f_{\rm GDA}(y,z,t|\eta)\,,
\\
\label{f_{GPD}(y,z,t|eta)}
f_{\rm GPD}(y,z,t|\eta) &\!\!\!=\!\!\!& \frac{-1}{2\pi^2}\frac{\partial}{\partial y} {\rm PV}\!\!\int_{-\eta}^1\!dx\frac{2(x-y)}{(x-y)^2-\eta^2 z^2}\;
 F(x,\eta,t)\,,
\\
\label{f_{GDA}(y,z,t|eta)}
f_{\rm GDA}(y,z,t|\eta) &\!\!\!=\!\!\!& \frac{-1}{2\pi^2} \frac{\partial}{\partial\eta y}{\rm PV}\!\!\int_{-1}^1\!dx \frac{2(x-\eta y)}{(x-\eta y)^2-z^2}\; \mathrm{F}(x,\eta,t)\,,
\end{eqnarray}
where the GPD and GDA contributions are separated.
Note that the inversion formula
\begin{equation}
\label{inversion}
f_{\rm GDA}(y,z,t|\eta) = 1/\eta^2 f(y,z,t|1/\eta)
\end{equation}
holds true, where $f(y,z,t|\eta)$ might be considered as extension of $f_{\rm GPD}(y,z,t|\eta)$ into the region $\eta>1$.

In the case that the GPD/GDA is continuous on the cross-over line, we can utilize partial integration  to express the derivation
w.r.t.\ $y$ and $\eta y$ as one w.r.t.\ $x$. Hence, the function (\ref{f(y,z,t|eta)}) can be written as
\begin{eqnarray}
\label{invRadon}
f(y,z,t)&\!\!\!=\!\!\!&\int_0^1\!\!\!d\eta\, f(y,z,t|\eta)\,,
\\
f(y,z,t|\eta) &\!\!\!=\!\!\!& \frac{-1}{2\pi^2} {\rm PV}\!\!\int_{-\eta}^1\!dx\left[\frac{2(x-y)}{(x-y)^2-\eta^2 z^2}\;
 \GPDf^\prime(x,\eta,t) - \frac{2(x+\eta y)}{(x+\eta y)^2-z^2}\;  \GDAf^\prime(-x,\eta,t) \right],
\nonumber \\
 &&\!\!\!\! + \frac{-1}{2\pi^2} {\rm PV}\!\!\int_{\eta}^1\!dx\left[\frac{2(x-y)}{(x-y)^2-\eta^2 z^2}\;
 \GPDf^\prime(x,-\eta,t) - \frac{2(x-\eta y)}{(x-\eta y)^2-z^2}\; \GDAf^\prime(-x,-\eta,t) \right],
  \nonumber
\end{eqnarray}
where $\GPDf^\prime(x,\eta,t)= \partial \GPDf(x,\eta,t)/\partial x$ and $\GDAf^\prime(x,\eta,t)= \partial \GDAf(x,\eta,t)/\partial x$.
Note that we have mapped the GDA integral over the interval $[-1,\eta]$ to an integral over the interval $[-\eta,1]$.

\subsubsection{Single integral inversion formulae}
\label{sec-single}

Based on the light--front wave function overlap representation, it has been shown  that for the Radon transform (\ref{F-D}), i.e., we have to set in all corresponding formulae of Ref.~\cite{Muller:2017wms} the formal spin parameter $s=0$, the DD can be restored
from the GPD $F^{\rm out}(x,\eta,t)= \GPDf(x,\eta,t)+\GPDf(x,-\eta,t)$ in the outer region by a {\em single} integral transformation in the complex plane,
\begin{equation}
\label{h(y,z,t)-3}
f(y,z,t)= \frac{-1}{2\pi i}\frac{\partial}{\partial y}\int_{-i \infty}^{i \infty}\! d r
\,
\frac{1 - y + z}{(1 + z + r z)^2}\,
F^{\rm out}\!\left(\frac{y + r z}{1 + z + r z},\frac{r - (1 + r) y}{1 + z + r z},t\!\right),
\end{equation}
within the support $|z|\le 1-y $ and $0\le y\le 1-y$. This novel formula is rederived in Appendix \ref{appendix} for a general GPD that not necessarily possess a wave function overlap representation.

Assuming that the integrand in (\ref{h(y,z,t)-3}) has for $0\le z \le 1-y$ and $0\le y\le 1$ only a cut on the real positive axis and vanishes at infinity, we can close the integration path by surrounding the first and forth quadrant of the complex $r$--plane and pick so up the   discontinuity on the positive real axis. After a  variable transformation $\eta=\frac{r - (1 + r) y}{1 + z + r z}$ the result reads
\begin{equation}
\label{h(y,z,t)-4}
f(y,z,t)=\frac{-1}{\pi}\frac{\partial}{\partial y}\int_{\frac{y}{1 - z}}^{\frac{1-y}{z}}\! d\eta
\,
\Im {\rm m} F^{\rm out}\!\left(y + \eta  z,\eta,t\!\right),
\end{equation}
where the imaginary part, $\Im {\rm m} F^{\rm out}(x,\eta) = \frac{1}{2}\left[F^{\rm out}(x,\eta+i\epsilon)-F^{\rm out}(x,\eta-i\epsilon)\right]$, is antisymmetric in $\eta$ and appears in the central region  $x\le |\eta|$.  The $(x,\eta)$ arguments of the GPD run from the cross-over line $(y/(1-z),y/(1-z))$  to $(1,(1-y)/z)$. Note that we have in (\ref{h(y,z,t)-4}) neglected the integral region $\eta \in[-\frac{y}{1 + z}, \frac{y}{1 - z}]$ that belongs to the outer region in which the imaginary part is absent.

\section{Analytic and numerical examples}
\label{Sec-examples}

\subsection{Elementary example}
\label{subsec-elementary}
Let us first consider a simple example, in which the GPD/GDA
is build from the function
\begin{equation}
\GPDf(x,\eta) = \frac{1}{\eta} \frac{x+\eta}{1+\eta}\quad
\quad\mbox{or}\quad  \GDAf(x,\eta) \equiv \frac{1}{\eta}\,\GPDf\!\left(\!\frac{x}{\eta},\frac{1}{\eta}\!\right) =  \frac{1+x}{1+\eta}
\end{equation}
and they read explicitly
\begin{equation}
F(x,\eta) = \theta(x+\eta) \frac{1}{\eta} \frac{x+\eta}{1+\eta} - \theta(x-\eta)\frac{1}{\eta} \frac{x-\eta}{1-\eta}
\end{equation}
\begin{equation}
 \mathrm{F}(x,\eta) = \theta(\eta-x) \frac{1+x}{1+\eta} + \theta(x-\eta) \frac{1-x}{1-\eta}\,.
\end{equation}
The GPD vanishes at the boundary $x=-\eta$ and $x=1$, hence, it is as the GDA continuous. Furthermore, a possible $D$--term contribution is absent.

Let us first find the DD by means of the Fourier transformed  GPD (\ref{FTofGPD}), valid within the restriction $|\eta|\le 1$. The result
\begin{equation}
\mathfrak{F}^{-1}[F](\kappa,\eta)=\frac{e^{-i \kappa }-e^{i \eta \kappa }}{\eta  (1+\eta ) \kappa ^2}-\frac{e^{-i \kappa } -e^{-i\eta \kappa }}{\eta  (1-\eta ) \kappa ^2}.
\end{equation}
is holomorphic in $\eta$ and reads after analytic continuation within the variable $\lambda= \eta \kappa$ as following
\begin{equation}
\mathfrak{F}_2^{-1}[f](\kappa,\lambda) =\mathfrak{F}^{-1}[\F](\kappa,\lambda/\kappa)=\frac{e^{-i \kappa }-e^{i\lambda}}{\lambda  (\kappa +\lambda )}-\frac{e^{-i \kappa }-e^{-i \lambda }}{(\kappa -\lambda ) \lambda }.
\end{equation}
Finally, the two-dimensional Fourier transform (\ref{DDfromGPD1}) provides us the DD
\begin{equation}
\label{f(y,z)}
f(y,z) = 1
\end{equation}
within  the support $|z| \le 1-y$ and $0\le y \le 1$.

\begin{figure}[t]
\begin{center}
\includegraphics[width=16cm]{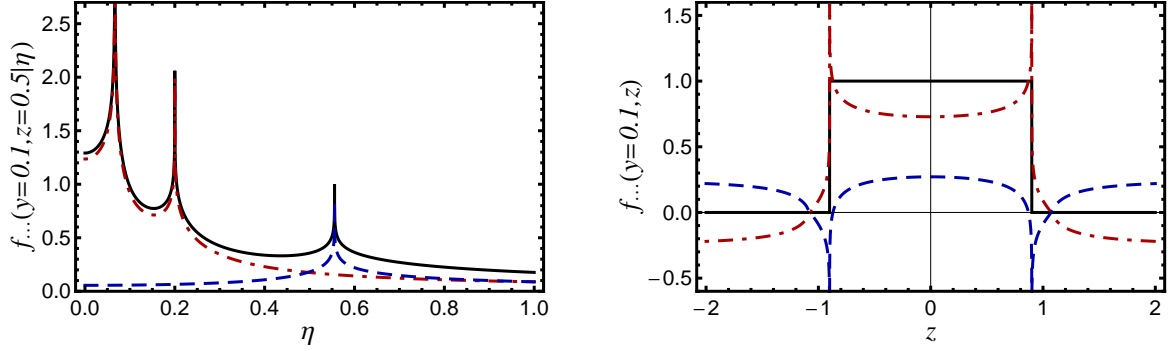}
\end{center}
\vspace{-5mm}
\caption{\small
Left: The function $f(y,z|\eta)$ (solid) and its GPD (dash--dotted) as well as GDA (dashed) components for $y=0.1$ and $z=0.5$ versus $\eta$.
Right: The resulting DD  $f(y,z)$ (solid), its GPD (dash--dotted) and GDA (dashed) components for $y=0.1$ versus $z$.
\label{Fig-fayz}
}
\end{figure}

As second method we consider the filtered backprojection. The differentiation of the extended GPD $\F$ w.r.t.~$x$ does not induce boundary terms. Hence, we can utilize formula (\ref{f(y,z,t|eta)}) with
\begin{equation}
 \GPDf^\prime(x,\eta) = \frac{1}{\eta}   \frac{1}{1+\eta} \quad\mbox{and}\quad  \GDAf^\prime(x,\eta) = \frac{1}{1+\eta}.
\end{equation}
We find for the GPD contribution
\begin{eqnarray}
\label{f_{GPD}}
f_{\rm GPD}(y,z|\eta) = \frac{-1}{\pi^2} \left\{\frac{\ln\left|\frac{(1-y)^2-z^2 \eta^2}{(\eta+y)^2-z^2 \eta^2}\right|}{\eta (1+\eta)} -
\frac{\ln\left|\frac{(1-y)^2-z^2 \eta^2}{(\eta-y)^2-z^2 \eta^2}\right|}{\eta (1-\eta)} \right\}
\end{eqnarray}
and for the GDA contribution
\begin{eqnarray}
\label{f_{GDA}}
f_{\rm GDA}(y,z|\eta) =\frac{-1}{\pi^2}\left\{\frac{\ln\left|\frac{\eta^2(1-y)^2-z^2 }{(1+\eta y)^2-z^2}\right|}{1+\eta} +
\frac{\ln\left|\frac{\eta^2(1-y)^2-z^2}{(1-\eta y)^2-z^2}\right|}{1-\eta} \right\}\,.
\end{eqnarray}
Both functions are related by inversion (\ref{inversion})
and possess only logarithmical singularities. They are located at
\begin{equation}
\eta \in \left\{\pm \frac{y}{1+z},\pm \frac{y}{1-z}, \pm \frac{z}{1-y},\pm \frac{1-z}{y},\pm \frac{1+z}{y} \right\},
\end{equation}
which are only relevant if they lie in the interval $\eta \in [0,1]$.
In the left panel of Fig.\ \ref{Fig-fayz} we display the functions (\ref{f_{GPD}}) (dash-dotted)  and (\ref{f_{GDA}}) (dashed) and their net result (solid) for $y=0.1$ and $z=0.5$ versus $\eta$.
The logarithmical singularities at $\frac{y}{1+z}=0.06\overline{6}$, $\frac{y}{1-z}=0.2$, and $\frac{z}{1-y}=0.5\overline{5}$ are clearly visible.

The integration over $\eta$ requires some care and the result can be put into the form
\begin{eqnarray}
f_{\rm GPD}(y,z) &\!\!\!= \!\!\!& \theta(y/(1-y))\theta((1-y)^2-z^2) -f_{\rm GDA}(y,z)\,,
\\
f_{\rm GDA}(y,z) &\!\!\!= \!\!\!&\frac{\theta((1-y)^2-z^2)}{4}- \frac{1}{2\pi ^2}\Re{\rm e} \left[\frac{1}{2}\ln^2\frac{1-y-z}{1-y+z}
+\text{Li}_2\!\!\left(\!\frac{-2 y}{1-y-z}\!\right)+\text{Li}_2\!\!\left(\!\frac{-2 y}{1-y+z}\!\right)\right].
\nonumber\\
\end{eqnarray}
Clearly, both functions possess logarithmical addenda, which are singular at $z=\pm (1-y)$ and have support in the whole $(y,z)$--plane.
In the net result they cancel each other and we recover the result (\ref{f(y,z)}).
This is illustrated in the right panel of Fig.\ \ref{Fig-fayz}.

Finally, let us employ the single integral inversion formulae (\ref{h(y,z,t)-3}) and (\ref{h(y,z,t)-4}) of Sec.\ \ref{sec-single}, where
$F^{\rm out}(x,\eta)=2(1-x)/(1-\eta^2)$.
Inserting
\begin{equation}
\frac{\partial}{\partial y}\frac{1 - y + z}{(1 + z + r z)^2}\,
F^{\rm out}\!\left(\frac{y + r z}{1 + z + r z},\frac{r - (1 + r) y}{1 + z + r z},t\!\right)=-\frac{2}{(1+r) (1+y+z-r (1-y-z))}
\end{equation}
into the inversion formula (\ref{h(y,z,t)-3}), closing the integration path by surrounding the first and forth quadrant, and utilizing Cauchy theorem immediately yields $f(y,z)=1$.
Alternatively, we can take the inversion formula (\ref{h(y,z,t)-4}), where
\begin{equation}
\frac{-1}{\pi}\frac{\partial}{\partial x}\Im{\rm m}F^{\rm out}(x,\eta+i \epsilon) = \delta(1-\eta) -  \delta(1+\eta).
\end{equation}
Since of support restrictions, only the term that is concentrated in $\eta=1$ contributes to
the integral on the r.h.s.~of (\ref{h(y,z,t)-4}).
Hence, we recover the already known result  $f(y,z)=1$.

\subsection{Photon GPD $\widetilde H$}

Let us now consider the quark helicity flip GPD $\widetilde H$  and the corresponding GDA for the photon, perturbatively calculated in LO accuracy of the QCD coupling constant at some renormalization point \cite{Friot:2006mm,ElBeiyad:2008ss}. Neglecting some overall factor, the defining functions read
\begin{eqnarray}
\label{GPDthGDAth}
\GPDth(x,\eta) =\frac{x+\eta}{2\eta} - \frac{1}{\eta} \frac{x+\eta}{1+\eta}
  &\Leftrightarrow&
\GDAth(x,\eta)  = \frac{1+x}{2\eta} - \frac{1+x}{1+\eta}.
\end{eqnarray}
They are represented in such a manner that the boundary conditions $\widetilde H(\eta,-\eta)=0$ and $\GDAtH(\pm 1,\eta)=0$ are satisfied.
 The first terms on the r.h.s.\ of these correspondences, denoted as
 \begin{eqnarray}
\delta\GPDth(x,\eta) =\frac{x+\eta}{2\eta}
  &\Leftrightarrow&
\delta\GDAth(x,\eta)  = \frac{1+x}{2\eta},
\end{eqnarray}
yield a constant GPD in the outer region $\delta\GPDth(x,\eta)+\delta\GPDth(x,-\eta)=1 $ and, consequently, it does not vanish at $x=1$. Furthermore, its $x$--dependent part belongs to an artificial $D$--term contribution
\begin{equation}
\label{tH-D}
d(x) = \frac{x}{2} - \frac{1}{2} {\rm sign}(x),
\end{equation}
calculated within the limiting procedure (\ref{d_F(x)-1}) or (\ref{d_F(x)-2}).
It lives in the central region and vanishes at the end-points, i.e., $d(x=\pm 1)=0.$ Since the full photon GPD
$\widetilde{H}^{(+)}(x,\eta)= \widetilde{H}(x,\eta) + \widetilde{H}(-x,\eta)$ is symmetric in $x$, this $D$--term does not contribute in the physical sector.

To perform the inverse Radon transform of $\delta {\widetilde H}(x,\eta)$, the $D$--term must be subtracted from the GPD, yielding a constant contribution that has the support $x \in [0,1]$
\begin{equation}
\delta {\widetilde H}(x,\eta) =\theta(x) \theta(1-x)\,.
\end{equation}

Clearly, its inverse Fourier transform (\ref{FTofGPD}) is independent on $\eta$, i.e., the two--dimensional Fourier transform (\ref{DDfromGPD1}) immediately provides us that the DD $\delta {\widetilde h}(y,z) = \delta(z)$  is concentrated in $z=0$.

In the filtered backprojection (\ref{F_i2f_i}) the derivative w.r.t.\ $x$ yields two $\delta$-functions $\partial \delta {\widetilde H}(x,\eta)/\partial x = \delta(x)-\delta(1-x)$.  Hence, this projection takes the form
\begin{equation}
\delta {\widetilde h}(y,z)=\int_0^\infty\!\!\!d\eta\, \delta {\widetilde h}(y,z|\eta)\,, \quad  \delta {\widetilde h}(y,z|\eta)= \frac{1}{2\pi^2} \left[\frac{y}{y^2-\eta^2 z^2}+\frac{1-y}{(1-y)^2-\eta^2 z^2}\right],
\end{equation}
which gives
\begin{equation}
\delta {\widetilde h}(y,z)=\frac{{\rm sign}(y)+{\rm sign}(1-y)}{2\pi^2}  \int_0^\infty\!\!\!d\eta\, \frac{1}{1-\eta^2 z^2}
= \theta(y(1-y)) \delta(z),
\end{equation}
where the integral represents $\pi^2 \delta(z)$.

This result follows also from the single integral formula (\ref{h(y,z,t)-3}). For a constant GPD we immediately read off the integral
\begin{equation}
\delta {\widetilde h}(y,z)= \frac{1}{2\pi i}\int_{-i \infty}^{i \infty}\! d r
\,
\frac{1}{(1 + z + r z)^2} \quad\mbox{for}\quad  0\le z \le 1-y
\end{equation}
which represents the correct answer $\delta(z)/2$. Symmetrization w.r.t.\ $z\to -z$ gives us again $\delta(z)$. Note that in the
dissipative inversion formula (\ref{h(y,z,t)-4}) this term is absent, since our assumption that the integral vanish at $r \to \infty$ does not hold true for $z=0$.

The second term on the r.h.s.\ of the correspondences (\ref{GPDthGDAth})  has been already treated in the previous section and leads to a constant DD $-1$. Finally, we can cast the DD ${\widetilde h}(y,z)$ into the form
\begin{equation}
\label{{widetilde h}(y,z)}
{\widetilde h}(y,z)= \delta(z)-1.
\end{equation}

\subsection{Photon GPD $H$}

Let us now consider the quark helicity conserved GPD $H$ and the corresponding GDA for the photon, independently calculated in \cite{Friot:2006mm} and \cite{ElBeiyad:2008ss}, respectively,
\begin{eqnarray}
\label{GPDGDAh(x,eta)}
\GPDh(x,\eta) = \frac{x+\eta}{2\eta} - \frac{x}{\eta} \frac{x+\eta}{1+\eta}
  &\Leftrightarrow& \GDAh(x,\eta) = \frac{1+x}{2 \eta }-\frac{x}{\eta }\frac{1+x}{1+\eta }.
\end{eqnarray}
Taking off the factor $x$ and $x/\eta$ in the second term on the r.h.s.\ of these correspondences provides with the results (\ref{f(y,z)}), (\ref{tH-D}), and (\ref{{widetilde h}(y,z)}) from the two previous sections  the following DD representation
\begin{equation}
\label{H(x,eta)-DD}
H(x,\eta) =  \int_0^1\!\!dy\!\!\! \int^{1-y}_{-1+y}\!\! dz\, \delta(x-y-\eta\, z)\left[ h(y,z) +\eta z \, \delta h(y,z) \right] +\theta(\eta^2-x^2) d(x/\eta)
\end{equation}
with $h(y,z) = \delta(z)-y$, $\delta h(y,z) = -1$, and $d(x)=x/2- {\rm sign}(x)/2$. One might now employ a `gauge' transformation
\begin{eqnarray}
\label{F-DD-1}
h(y,z) \to 
h(y,z) - \frac{\partial}{\partial z} z \int_y^{1-|z|}\!\!dw\, \delta h(w,z)\,, \;\;
d(x)\to d(x) - x \int_0^{1-|x|}\!\!dw\, \delta h(w,x)
\quad
\end{eqnarray}
to bring this Radon transform (\ref{H(x,eta)-DD}) into the standard form (\ref{F-D}). Evaluating the integrals yields the final result
\begin{eqnarray}
\label{h(y,z)-photon-result}
h(y,z)=1-2y -2|z| + \delta(z)  \quad\mbox{and}\quad d(x)= -\frac{x}{2}(1-2|x|)- \frac{1}{2}{\rm sign}(x)\,.
\end{eqnarray}

\begin{figure}
\begin{center}
\includegraphics[width=0.5\linewidth]{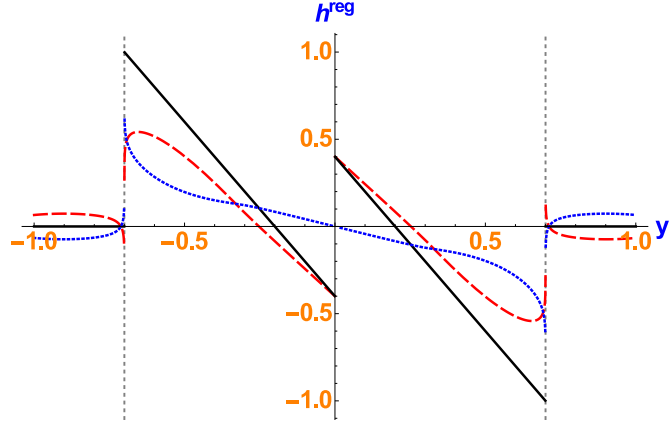}
\end{center}
\caption[]{\small The functions $h_{\rm GPD}^{\rm reg}(y,z)$  (dashed), $h_{\rm GDA}^{\rm reg}(y,z)$ (dotted), and the net result $h^{\rm reg}(y,z)$ (solid) for $z = 0.3$ versus $y$.}\label{F3_gpd_reg_sum}
\end{figure}

Alternatively, we might employ the filtered backprojection (\ref{f(y,z,t|eta)}). The regular DD part arises from the second term on the
r.h.s.\ of the correspondences (\ref{GPDGDAh(x,eta)}),
\begin{eqnarray}
\label{GPDGDAhreg(x,eta)}
\GPDh^{\rm reg}(x,\eta) =  - \frac{x}{\eta} \frac{x+\eta}{1+\eta}
  &\Leftrightarrow& \GDAh^{\rm reg}(x,\eta) = -\frac{x}{\eta }\frac{1+x}{1+\eta },
\end{eqnarray}
where the corresponding $D$--term contribution, given in terms of
\begin{eqnarray}
\label{GPDGDAhregD(x,eta)}
d^{\rm reg}(x)= -x (1-|x|)\,,
\end{eqnarray}
must be subtracted. Tactically, we consider in the following the full photon GPD $H^{\rm reg}(x,\eta)$ [DD $h^{\rm reg}(y,z)$] that is antisymmetrized in $x$ [$y$]. After integration over $x$, the GPD and GDA contributions are rather lengthy and they can be written as
\begin{eqnarray}
  &&h_{\rm GPD}^{\rm reg} (y,z|\eta) =
 \nonumber\\
 &&-\frac{1}{2\pi^2}\left\{
  \frac{1}{1-\eta^2}\left(4(y+z\eta)\ln\left|\frac{1-(y+z\eta)^2}{\eta^2-(y+z\eta)^2}\right| -
  2\ln\left|\frac{(1-y-z\eta)(\eta+y+z\eta)}{(1+y+z\eta)(\eta-y-z\eta)}\right|\right)-
  \right.
 \nonumber\\
  &&
  \left. 
  \frac{4}{\eta^2}(y+z\eta)\ln\left|\frac{\eta^2-(y+z\eta)^2}{(y+z\eta)^2}\right| +
  \frac{2}{\eta(1+\eta)}\ln\left|\frac{\eta-y-z\eta}{\eta+y+z\eta}\right| 
+
 \left\{\eta\rightarrow -\eta\right\}\right\}
\end{eqnarray}
and
\begin{eqnarray}
&& h_{\rm GDA}^{\rm reg} (y,z|\eta) =
 \nonumber\\
&& -\frac{1}{2\pi^2}\left\{
  \frac{\eta}{1-\eta^2}\left(4(y\eta+z)\ln\left|\frac{1-(y\eta+z)^2}{\eta^2-(y\eta+z)^2}\right|-
  2\ln\left|\frac{(1-y\eta-z)(\eta+y\eta+z)}{(1+y\eta+z)(\eta-y\eta-z)}\right|\right)
  \right. -
  \nonumber\\
&&  \left.
  \frac{4}{\eta}(y\eta+z)\ln\left|\frac{\eta^2-(y\eta+z)^2}{(y\eta+z)^2}\right| +
  \frac{2}{1+\eta}\ln\left|\frac{\eta-(y\eta+z)}{\eta+y\eta+z}\right|
+ \left\{\eta\rightarrow -\eta\right\} \right\}.
\end{eqnarray}
Note that they satisfy the inversion formula (\ref{inversion}) and that they contain logarithmical singularities, which are integrable.
Subsequent integration over $\eta$ gives the regular part  $h^{\rm reg}= h^{\rm reg}_{\rm GPD} + h^{\rm reg}_{\rm GDA} $:
\begin{eqnarray}\label{F1_dd}
 h^{\rm reg}(y,z)&\!\!\!=\!\!\!& (1-2|y|-2|z|){\rm sign}(y) \theta((1-|y|)^2-z^2)\theta(1-|y|)
\nonumber\\
 h^{\rm reg}_{\rm GDA} (y,z) &\!\!\!=\!\!\!&  h^{\rm reg}(y,z)-h_{\rm GPD}^{\rm reg}(y,z) 
 ,
\end{eqnarray}
where $h_{\rm GPD}^{\rm reg}(y,z)$ is a cumbersome function, which  is given in  appendix \ref{appendix2}.
The functions $h_{\rm GPD}^{\rm reg}(y,z)$  (dashed), $h_{\rm GDA}^{\rm reg}(y,z)$ (dotted) and the net result $h^{\rm reg}(y,z)$ (solid)  are shown in Fig.\ \ref{F3_gpd_reg_sum} for $z = 0.3$ versus $y$. As before in our elementary example, see Sec.\ \ref{subsec-elementary}, the GPD and GDA components possess no support restriction, while the DD $h^{\rm reg}(y,z)$ has the well--known support restriction, given by a rhombus $|z| \le 1-|y|$ and $|y|\le 1$. In addition, the GPD and GDA components are now finite at the boundary $y=\pm (1-|z|)$, however, they are discontinuous, compare with Fig.\ \ref{Fig-fayz}.

Let us finally mention that the DD $h(y,z)$ can be also straightforwardly obtained by means of Fourier transform, see Sec.\ \ref{Sec-Fourier transform}, or by means of single integral inversion formulae, see Sec.\ \ref{sec-single}. In particular, the transform (\ref{h(y,z,t)-3}),
which requires the analytic continuation of the GPD $H(x,\eta)$ in the outer region,
can be trivially performed by utilizing Cauchy theorem, which immediately leads to the result of $h(y,z)$, given in (\ref{h(y,z)-photon-result}). The corresponding $D$--term contribution might be obtained by the limiting procedure (\ref{d_F(x)-1}) or (\ref{d_F(x)-2})
and its negative sign \cite{Gabdrakhmanov:2012aa} is compatible with stability of a virtual quark cloud in the photon.

\subsection{Robustness of inverse Radon transforms}
To study the robustness of the various versions of inverse Radon transforms let us consider the RDDA for the DD-function at $t=0$ \cite{Radyushkin:2013hca},
\begin{equation}
\label{RDDA}
f(y,z) = \frac{\Gamma(2-\alpha +\beta)}{\Gamma(1-\alpha)\Gamma(1+\beta)}\, y^{-\alpha} (1-y)^{\beta-1}  \frac{\Gamma\!\left(\frac{3}{2}+b\right) }{\Gamma\!\left(\frac{1}{2}\right) \Gamma(1+b)} \left[1- \frac{z^2}{(1-y)^2}\right]^b.
\end{equation}
Here, $\frac{\Gamma(2-\alpha +\beta)}{\Gamma(1-\alpha)\Gamma(1+\beta)}\, y^{-\alpha} (1-y)^{\beta} $ is the parton distribution function, which lowest $x$--moment is normalized to one.
In particular, an integer $b$ value allows us to express the GPD in terms of rational functions and an incomplete ${\rm B}$--function,
\begin{equation}
{\rm B}(z,\alpha,\beta)=\int_0^z\!dt\, t^{\alpha-1}(1-t)^{\beta-1}.
 \end{equation} Moreover, for integer $\beta$ values the defining GPD function reduces to a rational function, e.g., for $\beta=3$
\begin{equation}
\label{RDDA-example}
\GPDf(x,\eta|\alpha,\beta=3,b=1)=\frac{4-\alpha }{4 \eta ^3}\left((1-x) (2-\alpha ) \eta +\eta ^2-x\right)\left(\frac{x+\eta }{1+\eta }\right)^{2-\alpha }.
\end{equation}
This defining function has besides poles at $\eta=0$ a discontinuity for $\frac{x+\eta }{1+\eta }\le 0$ while for our choice of integer $\beta$ a possible cut
for $\frac{x+\eta }{1+\eta }\ge 1$ is absent.

The elementary GPD, which we considered in Sec.\ \ref{subsec-elementary}, follows from the RDDA by setting
$\alpha=0$, $\beta=1$, and $b=0$. It is worth to mention that it possesses an overlap representation,
\begin{equation}
\GPDf(x,\eta)+\GPDf(x,-\eta)\propto\frac{1}{1-x} \frac{1-x}{1+\eta}\frac{1-x}{1-\eta}
\end{equation}
which saturates in the outer region $x \ge \eta$  the positivity bound \cite{Pobylitsa:2002iu}
\begin{equation}
\left|F(x,\eta,t)\right| \le \sqrt{F\!\left(\!\frac{x+\eta}{1+\eta},\eta=0,t=0\!\right) F\!\left(\!\frac{x-\eta}{1-\eta},\eta=0,t=0\!\right)/(1-\eta^2)}.
\end{equation}
For a generic proton GPD, where we set in the following $\alpha=1/2$, $\beta =3.3$, and $b=1$,
this positivity bound is violated. Hence, such a GPD possesses no overlap-representation and, thus,  it is {\em not} separable in the variables $x^{\rm i}=\frac{x+\eta}{1+\eta}$ and $x^{\rm f}=\frac{x-\eta}{1-\eta}$.

In what follows we study the robustness and fastness of the numerics for the inverse Radon transform
by utilizing the elementary ($\alpha=0$, $\beta=1$, and $b=0$) and generic proton ($\alpha=1/2$, $\beta=3.3$, and $b=1$) GPDs. We compare three numerical methods for the inverse Radon transform:
\begin{itemize}
\item Filtered backprojection in the version of (\ref{invRadon}).
\item Single integral transform in the complex plane (\ref{h(y,z,t)-3}).
\item Single dissipative integral transform (\ref{h(y,z,t)-4}).
\end{itemize}

\subsubsection{Filtered backprojection}

\begin{figure}[t]
\begin{center}
\includegraphics[width=16cm]{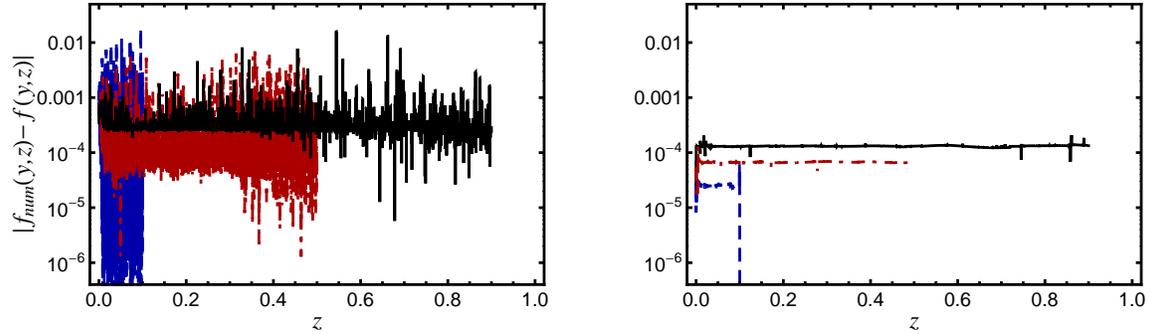}
\end{center}
\vspace{-5mm}
\caption{\small The absolute error $|f_{\rm num}(y,z)-f(y,z)|$ of the numerically evaluated filtered backprojection (\ref{invRadon}) from the analytic form of the DD (\ref{RDDA}) for
$y=0.1$ (solid), $y=0.5$ (dash-dotted), and $y=0.9$ (dashed) versus $z$.
Left: Elementary example $\alpha=0$, $\beta=1$, and $b=0$.
Right: Generic proton GPD $\alpha=1/2$, $\beta=3.3$, and $b=1$.
\label{Fig-accuracyFBP}
}
\end{figure}
We explore now the filtered backprojection, by means of formula (\ref{invRadon}), where the GPD $\GPDf$ and GDA $\GDAf$ building blocks are given in analytic form. We calculated numerically the DD and compared it with its analytic form  (\ref{RDDA}) of the DD.  The principal value integrals are performed
with a subtraction procedure, where the subtracted term is analytically evaluated. The pole terms in $\GPDf$ at $\eta=0$, cf.\ equation (\ref{RDDA-example}), cancel each other, however, their numerical evaluation requires some care. Thus, we cut--out the region $\eta \in [0,\eta_0]$ and choose $\eta_0=10^{-4}$. The numerical integrations are done with the standard MATHEMATICA routine within a precision of $10^{-16}$ and requiring an accuracy of $10^{-4}$ in the final result.
In the left and right panel of Fig.\ \ref{Fig-accuracyFBP} we display
the absolute error, $|f_{\rm num}(y,z)-f(y,z)|$, of the numerical and analytical (exact) result for the elementary GPD  example, see Sec.\ \ref{subsec-elementary}, by taking $\alpha=0,$ $\beta=1$, and $b=0$ and for generic proton GPD parameters $\alpha=1/2$, $\beta=3.3$, and $b=1$, respectively. As one realizes the elementary example is numerically more challenging than the generic proton one. In the former case, the error is rather noisy and can increases to the $10^{-2}$ level, while in the latter case the error is on the $10^{-4}$ level or even below.

\begin{figure}[t]
\begin{center}
\includegraphics[width=16cm]{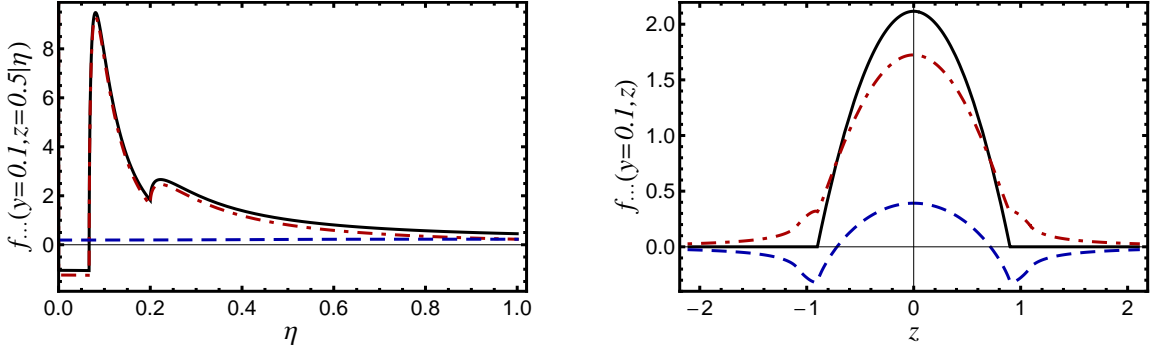}
\end{center}
\vspace{-5mm}
\caption{\small
Left: The function $f(y,z|\eta)$ (solid) for $\alpha=1/2$, $\beta=3.3$ and $b=1$ as well as  its GPD (dash--dotted) and  GDA (dashed) components for $y=0.1$ and $z=0.5$ versus $\eta$.
Right: The resulting DD  $f(y,z)$ (solid), its GPD (dash--dotted) and GDA (dashed) components for $y=0.1$ versus $z$.
\label{Fig-RDDAyz}
}
\end{figure}
The reason for these rather different numerical features is the behavior of the DD at the boundary $z=\pm (1-y)$.
In the left panel of Fig.\ \ref{Fig-RDDAyz} we plot the corresponding DD $f(y,z|\eta)$ (solid) and its GPD (dash-dotted) as well as its GDA (dashed) component for $y=0.1$ and $z=0.5$. Compared to our elementary example, shown in Fig.\ \ref{Fig-fayz}, the logarithmical singularities are washed out. In the right panel, where we show the resulting DD (solid) and its components, we also realize that the singularities at the boundary $z=\pm (1-y)$ disappear in the GPD (dash-dotted) and GDA (dashed) components.

\subsubsection{Single integral transform in the complex plane}

\begin{figure}[t]
\begin{center}
\includegraphics[width=16cm]{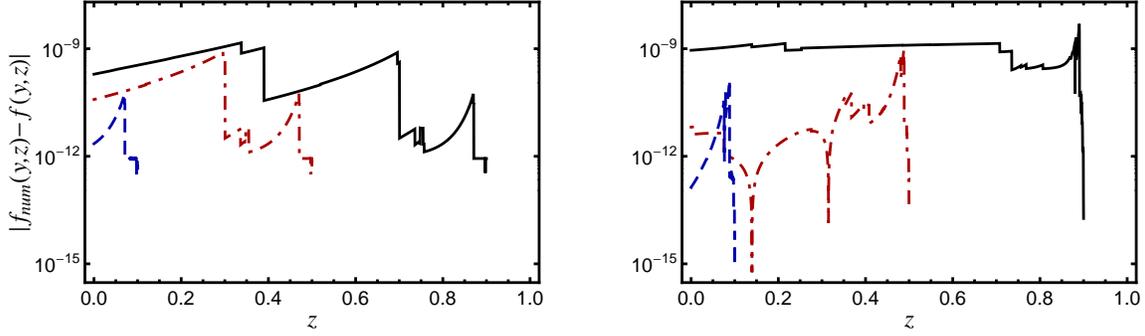}
\end{center}
\vspace{-5mm}
\caption{\small The absolute error $|f_{\rm num}(y,z)-f(y,z)|$ of the numerically evaluated single integral transform (\ref{h(y,z,t)-3}) from the analytic form of the DD for
$y=0.1$ (solid), $y=0.5$ (dash-dotted), and $y=0.9$ (dashed) versus $z$.
Left: Elementary GPD example ($\alpha=0$, $\beta=1$, and $b=0$).
Right: Generic proton GPD ($\alpha=1/2$, $\beta=3.3$, and $b=1$).
\label{Fig-accuracyComInt}
}
\end{figure}

For the test of the single integral transform (\ref{h(y,z,t)-3}) within the RDDA, the working precision for the evaluation of the integral was again taken to be $10^{-16}$. The results are plotted in Fig.\ \ref{Fig-accuracyComInt} and as one realizes the absolute error in both cases, the elementary example and the generic proton GPD is around $10^{-9}$ or even below this value. Compared  to the filtered backprojection, displayed in Fig.\ \ref{Fig-accuracyFBP}, this is an improvement by at least five orders of magnitude. We checked that an increase of the working precision decreases the error. It is clear that the evaluation
of a single integral is much faster than the filtered backprojection. Requiring an accuracy of $10^{-4}$ the single integral transform is compared to the filtered backprojection in our implementation roughly a factor 50 faster.

We also checked that for a selection of non-integer $\alpha,\beta,$ and $b$ parameters  the original DD is reproduced within a numerical accuracy of $10^{-9}$.
Thereby, the extension of the GPD in the complex plane was evaluated numerically by means of the parameter integral
\begin{equation}
F^{\rm out}(x,\eta,t)= \frac{2 (1-x)}{1-\eta ^2}\int_0^1\!dw\, f\!\left(\!\frac{x+\eta}{1+\eta} w + \frac{x-\eta}{1-\eta} (1-w),\frac{1-x }{1-\eta ^2}(1+\eta -2 w),t\!\right),
\end{equation}
which straightforwardly follows from the integral representation (\ref{GPDf(x,eta,t}) of the defining function $\GPDf$.
Due to the double integral the numerics is rather slow, however, remains very accurate.

\subsubsection{Single dissipative integral transform}

It is also possible to employ the dissipative integral transform (\ref{h(y,z,t)-4}), however, as we have seen in our elementary example in
Sec.~\ref{subsec-elementary} it might contain numerically non-integrable singularities in form of generalized mathematical functions, e.g., Dirac`s $\delta$-function. This is also the case for RDDA where the singularities have a $|1-\eta|^{\alpha-\beta}$ form. Such generalized mathematical functions  with $\beta-\alpha > 1$ can be  integrated by analytic regularization \cite{GelShi64}. Note that for integer $\beta-\alpha$ these generalized mathematical function provide derivatives of the $\delta$-function. Knowing the analytic form of the imaginary part, we can define a regularized function, e.g., for the $b=0$ case, it takes the form
\begin{equation}
\Im {\rm m} F^{\rm out}\!\left(x,\eta,t\!\right)= \theta(1-\eta) (1-\eta)^{\alpha-\beta} \Im {\rm m} F_-^{\rm reg}\!\left(x,\eta,t\!\right)+
\theta(\eta-1) (\eta-1)^{\alpha-\beta} \Im {\rm m} F_+^{\rm reg}\!\left(x,\eta,t\!\right).
\end{equation}
The DD reads then as follows
\begin{eqnarray}
f(y,z)&\!\!\!=\!\!\!&\frac{-1}{\pi}\int_{\frac{y}{1 - z}}^{1}\! d\eta
\left[\frac{1}{1-\eta}\right]^{(n)}_+ \Im {\rm m} F_-^{\prime{\rm reg}}(y+\eta z,\eta)
+\frac{-1}{\pi}\int_{1}^{\frac{1-y}{z}}\! d\eta
\left[\frac{1}{\eta-1}\right]^{(n)}_+ \Im {\rm m} F_+^{\prime{\rm reg}}(y+\eta z,\eta)
\nonumber\\
&&+\frac{-1}{\pi}
\sum_{m=0}^{n} \frac{(-1)^m\left(\frac{1-y-z}{1-z}\right)^{1+m+\alpha -\beta }
\Im {\rm m} F^{\prime{\rm reg}(m)}_-(y+z,1)}{(1+m+\alpha -\beta)m! }
\nonumber\\
&&+\frac{-1}{\pi}
\sum_{m=0}^{n} \frac{\left(\frac{1-y-z}{z}\right)^{1+m+\alpha -\beta }
\Im {\rm m} F^{\prime{\rm reg}(m)}_+(y+z,1)}{(1+m+\alpha -\beta)m! },
\end{eqnarray}
where $n=[\beta-\alpha-1]$ is the integer part of $\beta-\alpha-1$, the $+$--definition, acting on a test function $f(\eta)$, is defined as
\begin{equation}
\left[\frac{1}{1-\eta}\right]^{(n)}_+ f(\eta) =  f(\eta) - \sum_{m=0}^n (\eta-1)^m \frac{1}{m!}\frac{d^m f(\eta)}{d\eta^m}\Big|_{\eta=1}.
\end{equation}

\begin{figure}[t]
\begin{center}
\includegraphics[width=14cm]{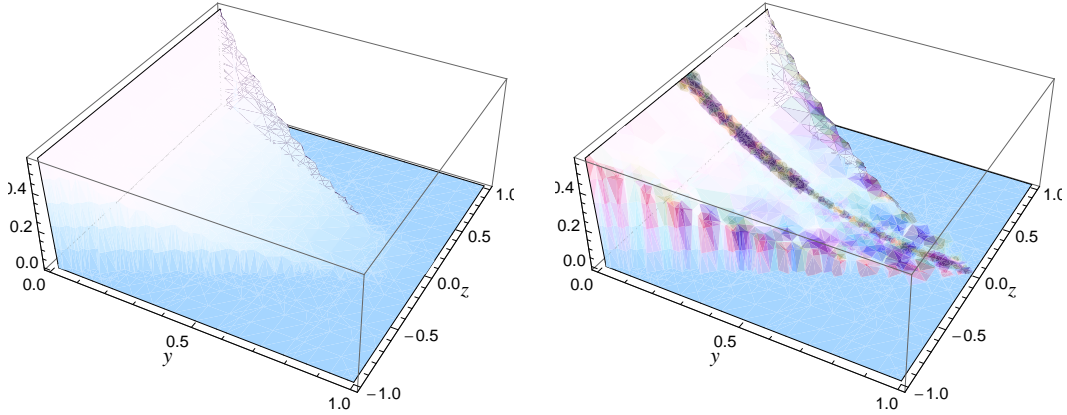}
\end{center}
\vspace{-5mm}
\caption{\small  RDDA for $\alpha=1/2$, $\beta=3.3,$ and $b=0$, multiplied with $\sqrt{y}$, versus $y$ and $z$ as a result of the single integral transform in the complex $r$--plane (left) and along the real $\eta$--axis (right).
\label{Fig-3Dplot}
}
\end{figure}
The complexity of such formulae increases with growing (integer) $b$--parameter. Since we do not have analytic expressions for non-integer $b$--parameter at hand, this dissipative method cannot be applied in such a case. Compared to the single integral transform (\ref{h(y,z,t)-3}), the dissipative based method might be faster, however, it is numerically challenging in the vicinity of $z=0$ and $z=1-y$. This is illustrated for $\alpha=1/2$, $\beta=3.3,$ and $b=0$ case in Fig.\ \ref{Fig-3Dplot}, where the numerical noise at the boundary and in the vicinity of $z=0$ is clearly visible, see right panel.
Requiring an accuracy of $10^{-3}$, the evaluation time for the 3D plots is 38 s and 15 s with an Intel core i7 processor @ 2.4 GHz.

\section{Summary}
\label{Sec-Conclusions}

In this article, we considered various forms of the inverse Radon transform, which provides the DD from a given (extended) GPD.
We gave both analytical and numerical examples.

For the filtered backprojection, which is a two--dimensional integral transform, on needs the extension of the GPD according to its support properties, which is an unique procedure. This  extension can be achieved in terms of the GDA, which follows from crossing. Knowing the defining GPD function $\GPDf(x,\eta,t)$ the GPD as well as the GDA can be obtained.
The exact inversion equation has been derived in terms of the defining function and applied to the photon leading order GPD as an analytic example and, numerically, to the RDDA model.

Notably the perturbative photon GPDs contain a $D$--term contribution and a piece that is entirely given by a skewless function, having the support $x\in [0,1]$. The later one corresponds  to a $\delta(z)$ singularity in the DD. Such a contribution is easily obtained in an analytic treatment, however, might be missed numerically.
In the physical sector the $D$--term vanishes for the quark helicity flip GPD $\widetilde H$, as it must be. For the quark helicity conserved GPD $H$ the odd moments of the $D$--term (\ref{h(y,z)-photon-result}) are negative.

Adopting a previous result, which was derived by means of the wave function overlap representation,
we showed that the inverse Radon transform could be also given in terms of single integrals along the imaginary axis or along the positive real axis. Here the analytic continuation of the GPD from the outer region into the complex plane, respectively, in the central region is needed, where in the latter case the imaginary part is essential. We stress that such an imaginary part arises only due to the analytic continuation and disappears if one takes the GPD support properties into account. We exemplified that these results also hold for GPDs that violate positivity bounds, i.e., do not possess an overlap representation.

Compared to the filtered backprojection, the single integral transforms allow for a simpler analytical treatment and an much faster numerical one. In the dissipative inversion formula one has to deal with generalized functions that require the use of analytical regularization. It can only be applied if the imaginary part of the GPD $F^{\rm out}$ is known in analytic form. In particular, the integral transform along the imaginary is straightforward to handle and it is both fast and numerically accurate.

\section*{Acknowledgements}
The authors would like to thank I.V.\ Anikin, A.V.\ Efremov, C. Lorc\'e, H.\ Moutarde, M.V.\ Polyakov and A.V.\ Radyushkin for useful
discussions and comments. I.G.\ is also indebted to A.N.\ Khairullin for helpful discussion.
D.M.\ is thankful for the support of Heisenberg-Landau  program of Germany-JINR collaboration and warm hospitality of the Bogoliubov Theoretical Laboratory,  where this article has been finalized.

\appendix
\section{Derivation of single integral transform}
\label{appendix}

The result (\ref{h(y,z,t)-3})  can be easily adopted also for GPDs that do not possess a wave function overlap representation. In analogy to the derivation in Sect.~3 of Ref.~\cite{Muller:2017wms} we represent the GPD in the outer region as a two--fold Laplace transform of a function $\Phi(\lambda^{\rm i},\lambda^{\rm f},t)$,
\begin{equation}
\label{F-overt}
F^{\rm out}(x,\eta,t) =F(x\ge \eta,\eta,t) = \frac{1}{1-x}
\int_0^\infty\! d\lambda^{\rm i} \int_0^\infty\! d\lambda^{\rm f} e^{-\frac{\lambda^{\rm i}}{1-x^{\rm i}}-\frac{\lambda^{\rm f}}{1-x^{\rm f}}}\, \Phi(\lambda^{\rm i},\lambda^{\rm f},t),
\end{equation}
where $x^{\rm i}= (x+\eta)/(1+\eta)$ and $x^{\rm f}= (x-\eta)/(1-\eta)$ are the momentum fractions of the struck parton in the initial and final  state. Plugging in the identity $\int^{\infty}_0\! d\lambda \delta(\lambda-\lambda^{\rm i}-\lambda^{\rm f})=1$ and due to  variable substitutions,
\begin{equation}
\lambda^{\rm i} = \frac{1-x-z(1-\eta)}{2(1-y)} \lambda\,, \quad \lambda^{\rm f} = \frac{1-x+z(1+\eta)}{2(1-y)} \lambda\,,
\end{equation}
 the representation (\ref{F-overt}) can be converted into the DD representation (\ref{F-D}).
In particular, we have
\begin{equation}
\frac{\theta(\lambda^{\rm i})\theta(\lambda^{\rm f})}{1-x}\delta(\lambda-\lambda^{\rm i}-\lambda^{\rm f}) d\lambda^{\rm i} d\lambda^{\rm f} =
\theta\!\left(\!\frac{y^{\rm i}\lambda}{1-y}\!\right)\theta\!\left(\!\frac{y^{\rm f}\lambda}{1-y}\!\right)\frac{\lambda{\rm Sign}(1-y)}{2(1-y)^2}  \delta(x-y-z\eta) dy dz
\end{equation}
with $y^{\rm i} = (1-y-z)/2$ and $y^{\rm f} = (1-y+z)/2$ as well as
$
e^{-\frac{\lambda^{\rm i}}{1-x^{\rm i}}-\frac{\lambda^{\rm f}}{1-x^{\rm f}}} = e^{-\lambda/(1-y)},
$
which provides us after rescaling of $\lambda \to \lambda (1-y)$ the radon transform (\ref{F-D}), where
the DD has the support $|z| \le (1-y)$ and reads as follows
\begin{equation}
\label{h(y,z)}
f(y,z,t)=  \frac{\theta(y^{\rm i})\theta(y^{\rm f})}{2}\int_0^\infty\! d\lambda\, \lambda\, e^{-\lambda}\, \Phi(y^{\rm i} \lambda, y^{\rm f} \lambda, t).
\end{equation}
In this equation we insert the two--dimensional inverse Laplace transform
of $(1-x)F^{\rm out}(x,\eta,t)$, see (\ref{F-overt}),
\begin{equation}
\Phi(y^{\rm i} \lambda, y^{\rm f} \lambda, t)= \frac{1}{(2\pi i)^2} \int_{c-i\infty}^{c+i\infty}\!\!
\!\int_{c-i\infty}^{c+i\infty}\!\!  dr^{\rm i}dr^{\rm f}\,  e^{y^{\rm i}\lambda  r^{\rm i}+y^{\rm f}\lambda  r^{\rm f}}
\frac{2F^{\rm out}\!\left(\!\frac{r^{\rm f}+r^{\rm i}-2}{ r^{\rm f}+r^{\rm i}},\frac{r^{\rm i}- r^{\rm f}}{ r^{\rm f}+r^{\rm i}},t\!\right)}{ r^{\rm f}+r^{\rm i}},
\end{equation}
 which provides us after a shift of the integration variables $r^{\rm i,f}\to r^{\rm i,f}+1$ and performing the integration over $\lambda$
\begin{equation}
\label{h(y,z)-1}
f(y,z,t)= \frac{\theta(y^{\rm i})\theta(y^{\rm f})}{2(2\pi i)^2}\frac{\partial}{\partial y}\int_{-i \infty}^{i \infty}\! d r^{\rm i}\!\!\int_{-i \infty}^{i \infty}\! d r^{\rm f}\;
\frac{
4F^{\rm out}\!\left(\!\frac{r^{\rm i}+r^{\rm f}}{2+r^{\rm i}+r^{\rm f}},\frac{r^{\rm i}-r^{\rm f}}{2+r^{\rm i}+r^{\rm f}},t\!\right)
}{(2+ r^{\rm i}+ r^{\rm f})^2
(y^{\rm i} r^{\rm i} +y^{\rm f}r^{\rm f}-y)
}\,.
\end{equation}
Here we assumed that $F^{\rm out}\!\left(\!\frac{r^{\rm i}+r^{\rm f}}{2+r^{\rm i}+r^{\rm f}},\frac{r^{\rm i}-r^{\rm f}}{2+r^{\rm i}+r^{\rm f}},t\!\right)/(2+ r^{\rm i}+ r^{\rm f})^2$ has only singularities on the l.h.s.\ of the integration paths. Furthermore, if it vanishes in the limit $r^{\rm f}\to \infty$, we can close the integration path w.r.t.\ $r^{\rm f}$--integration by surrounding the first and fourth quadrant and pick so up the pole at $r^{\rm f}=(y-  y^{\rm i} r^{\rm i})/ y^{\rm f}$, which yields (\ref{h(y,z,t)-3}). Note that the pole contribution
does not contribute for $y<0$. Consequently, we have recovered the support  of the DD, given by $|z|\le 1-y$  and $0 \le y \le 1$.

\section{Result for $h_{\rm GPD}^{\rm reg}(y,z)$}
\label{appendix2}

\begin{eqnarray}
\label{F1_gpd_reg}
h_{\rm GPD}^{\rm reg}(y,z) &\!\!\!=\!\!\!& -\frac{1}{4\pi^2}\Re{\rm e}\Bigg[
\Bigg\{(1+2y)\left(\ln^2\frac{1+y-z}{1+y+z}-\ln^2\frac{-1+y-z}{1+y+z}\right)+
4\text{Li}_2\!\left(\frac{2-2z}{1+y-z}\right)
\nonumber\\
&&\phantom{ -\frac{1}{4\pi^2}\Re{\rm e}\Bigg\{} - \left\{{ y\atop z}\; \rightarrow\; { -y\atop -z}\right\}\Bigg\}
\\
&&\phantom{ -\frac{1}{4\pi^2}\Re{}} +\Bigg\{
4z \text{Li}_2\!\left(\frac{2+2z}{1+y+z}\right)-4z \text{Li}_2\!\left(\frac{2z}{1+y+z}\right)-4(1+2z) \text{Li}_2\!\left(\frac{1+z}{y}\right)
\nonumber\\
&&\phantom{ -\frac{1}{4\pi^2}\Re{\rm e}\Bigg\{}
+8z \text{Li}_2\!\left(\frac{z}{y}\right)+8(1+y+z) \ln(1+y+z)-8(y+z)\ln(y+z)
\nonumber\\
&&\phantom{ -\frac{1}{4\pi^2}\Re{\rm e}\Bigg\{} +\left\{{ y\atop z}\; \rightarrow\; { y\atop -z}\right\}
- \left\{{ y\atop z}\; \rightarrow\; { -y\atop z}\right\}- \left\{{ y\atop z}\; \rightarrow\; { -y\atop -z}\right\}\Bigg\}
\Bigg].\nonumber
\end{eqnarray}


\end{document}